\documentclass[twocolumn,amsmath,amssymb,nofootinbib,eqsecnum,tighten]{revtex4}

\usepackage{graphicx}
\usepackage{dcolumn} 
\usepackage{bm,bbm} 
\usepackage{curves} 
\usepackage{epic}
\usepackage{amsthm}
\usepackage[mathcal]{euscript}
\usepackage{epsf,times}

\def\=d{\, {\buildrel \rm def  \over =} \,}

\def\sqr#1#2{{\vcenter{\vbox{\hrule height.#2pt \hbox{\vrule width.#2pt height#1pt \kern#1pt \vrule width.#2pt}\hrule height.#2pt}}}}

\def\beq#1{\begin{equation} \label{#1}}
\def\eeq{\end{equation}}
\def\ben{\begin{equation*}}
\def\een{\end{equation*}}
\def\bequa{\begin{eqnarray}}
\def\eequa{\end{eqnarray}}
\def\re#1{(\ref{#1})}

\def\bf#1{\bm{#1}}

\def\path{figures}

\begin{document}

\title{Screening in $(d+s)$-wave superconductors: Application
to Raman scattering}

\author{Andreas P. Schnyder}
\affiliation{
Condensed Matter Theory Group, Paul Scherrer Institute, CH-5232 Villigen PSI, Switzerland 
            }
\author{Christopher Mudry}
\affiliation{
Condensed Matter Theory Group, Paul Scherrer Institute, CH-5232 Villigen PSI, Switzerland 
            }
            \author{Dirk Manske}
\affiliation{
Institut f\"ur Theoretische Physik, ETH Z\"urich, H\"onggerberg, CH-8093 Z\"urich, Switzerland
            }
\affiliation{
Max-Planck-Institut f\"ur Festk\"orperforschung, D-70569 Stuttgart, Germany
            }

\date{\today}

\begin{abstract}
  We study the polarization-dependent electronic Raman response of
  untwinned YBa$_2$Cu$_3$O$_{7-\delta}$ superconductors 
  employing a tight-binding band
  structure with anisotropic hopping matrix parameters and a 
  superconducting gap with a mixing of $d$- and $s$-wave symmetries.
  Using general arguments we find   screening terms in the $B^{\ }_{1g}$
  scattering channel 
  which are required by gauge invariance. 
  As a result, we obtain a small but measurable
  \emph{softening} of the pair-breaking peak, whose position has been
  attributed for a long time to twice the superconducting gap maximum.  
  Furthermore, we predict superconductivity-induced changes in the
  phonon line shapes that could provide a way to detect the isotropic $s$-wave
  admixture to the superconducting gap.
\end{abstract}

\maketitle

\section{Introduction}
The symmetry of the superconducting (SC) order parameter 
in cuprate high-$T_c$ superconductors is now agreed to be unconventional 
after early intense theoretical and experimental debates.\cite{Mueller95}
Phase-sensitive experiments such as
corner-junction superconducting quantum interference device
experiments \cite{Wollmann93} and tricrystal
experiments on YBa$_2$Cu$_3$O$_{7-\delta}$ (YBCO) (Ref.~\cite{Tsuei94})
show that the SC order parameter undergoes sign changes 
in the Brillouin zone (BZ) that are consistent with the $d_{x^2-y^2}$ symmetry.
A sign change of a SC order parameter with the $d_{x^2-y^2}$ symmetry
implies the existence of four nodal points in the Brillouin zone
that have been observed using momentum-resolved probes such as
inelastic neutron scattering (INS) or 
angle-resolved photoemission spectroscopy (ARPES).
Correspondingly, the simple gap
\begin{eqnarray}
\Delta^{\ }_{\boldsymbol{k}}= 
\Delta^{\ }_0(\cos k^{\ }_x-\cos k^{\ }_y)/2 
\label{eq: def pure d-wave gap}
\end{eqnarray}
with a $d_{x^2-y^2}$ symmetry on the Brillouin zone
has often been used as a starting point of a quantitative interpretation of 
these experiments. 

Of course, a sign change in the SC gap does not preclude 
a gap that is more complicated than the simple $d_{x^2-y^2}$ gap
(\ref{eq: def pure d-wave gap}). For those high-$T_c$ cuprates
with a tetragonal crystalline structure, higher $d$-wave harmonics
have been invoked to explain ARPES measurements of the 
magnitude of the SC gap.\cite{Mesot99}
For the cuprate family YBa$_2$Cu$_3$O$_{7-\delta}$, 
which exhibits quite strong orthorhombic distortions,
one expects corrections to the $d_{x^2-y^2}$ gap
(\ref{eq: def pure d-wave gap}) on symmetry grounds alone.
This expectation has been confirmed by several experimental methods such as
ARPES studies,\cite{Lu01} 
INS studies,\cite{Hinkov04}
and measurements of Josephson currents,\cite{Smilde05,Hilgenkamp06}
and has also been investigated theoretically.\cite{beal-monod-and-co,Nemetschek98,einzel99}

Polarization-dependent electronic Raman scattering 
also probes the momentum dependence of the magnitude of the
superconducting order parameter and has provided
yet one more piece of evidence for the 
$d_{x^2-y^2}$-wave pairing scenario.\cite{Chen93,Opel00,Devereaux07}  
In particular, for tetragonal high-$T_c$ cuprates, 
one finds (a) various low-energy power laws in different polarization channels
that are consistent with the existence of nodal points for the gap
and (b) the pair-breaking peak in the $B^{\ }_{1g}$ channel at energy twice 
the superconducting gap maximum $\widetilde{\Delta}^{\ }_{0}$
seen by other means.\cite{Devereaux95}

In this paper, we are going to investigate the consequences for
polarization-dependent Raman scattering 
of a subdominant admixture of an isotropic $s$-wave component to 
the gap~(\ref{eq: def pure d-wave gap})
which should be of relevance to orthorhombic 
high-$T_{c}$ superconductors of the YBCO family.
Assuming the existence of well-defined SC quasiparticles,
we compute the polarization-dependent electronic 
Raman-scattering cross section including the effects of
(i) an orthorhombic tight-binding dispersion with $(d+s)$ gap,
(ii) a long-range Coulomb interaction treated within the random-phase approximation (RPA),
(iii) and the effective mass approximation for the Raman vertex.
We show that the pair-breaking peak in the $B^{\ }_{1g}$ channel is softened 
by an amount proportional to the isotropic $s$ component to the gap in that
it occurs at a lower value than the absolute maximum of the gap
on the ``normal-state'' Fermi surface.
We also show that the $A^{\ }_{1g}$ channel develops a double peak structure
with the peak separation proportional to the isotropic $s$ component to the gap.
Furthermore, we compute superconductivity-induced changes
in the phonon line shapes, and argue how Raman scattering on phonons allows us to extract
a signature of a subdominant and isotropic $s$-wave component to the gap.

\section{Theory}
The differential cross section in a Raman-scattering experiment 
for a momentum transfer $\boldsymbol{q}$ 
that is small compared to the extension of the Brillouin zone
is proportional to 
$[1+n(\omega)]\,\chi^{\prime\prime}_{\gamma}(\omega)$,
where $n$ denotes the Bose distribution, 
$\omega$ the frequency of the incoming plane wave, 
and 
\begin{subequations}
\label{eq: master equations}
\begin{eqnarray}
\chi^{\ }_{\gamma}(\omega)=
\left(
\chi^{\prime}_{\gamma}
+
i
\chi^{\prime\prime}_{\gamma}
\right)(\omega)
\equiv
\chi^{\ }_{\gamma}(\boldsymbol{q}\approx0,\omega+i\eta)
\end{eqnarray}
is the linear-response function for the density operator 
\begin{eqnarray}
\rho^{\ }_{\bm{q}} =
\sum_{\bm{k}}
\sum_{\sigma=\uparrow,\downarrow} 
\gamma^{\ }_{\bm{k}} 
c^{\dag}_{\bm{k}+\bm{q},\sigma} c^{\ }_{\bm{k},\sigma}.
\end{eqnarray}
The coupling between the SC quasiparticles, linear superpositions of
the fermionic creation $c^{\dag}_{\bm{k},\sigma}$ and annihilation
$c^{\ }_{\bm{k},\sigma}$ operators, 
and the incoming (outgoing) plane wave with the polarization vector
$\hat{e}^{I}_{}$ ($\hat{e}^{O}_{}$)
is here approximated by
\begin{eqnarray}
\gamma^{\ }_{\bm{k}}\propto
\sum_{\alpha,\beta}
\hat{e}^{O}_{\alpha}
\frac{\partial^2\varepsilon^{\ }_{\bm{k}}}
{\partial k^{\ }_{\alpha}\partial k^{\ }_{\beta}}
\hat{e}^{I}_{\beta}
\label{eq: Raman vertex}
\end{eqnarray}
in the nonresonant limit.
The ``normal-state'' dispersion 
\begin{eqnarray}
\varepsilon^{\ }_{\bm{k}}&=&
- 
2t 
\left[
(1+\delta^{\ }_0)\cos k^{\ }_x 
+ 
(1-\delta^{\ }_0)\cos k^{\ }_y 
\right] 
\nonumber\\
&&
\qquad -
4t'\cos k^{\ }_x\cos k^{\ }_y 
- 
\mu
\label{eq: dispersion}
\end{eqnarray}
(see Figs. 1 and 2 in Ref. \onlinecite{Schnyder06})
combines with the gap
\begin{eqnarray}
\Delta^{\ }_{\bm{k}}=
\frac{\Delta^{\ }_0}{2} 
\left( 
\cos k^{\ }_x 
- 
\cos k^{\ }_y
\right) 
+ 
\Delta^{\ }_s
\label{eq: def isotropic s wave gap}
\end{eqnarray}
to give the SC quasiparticle dispersion
$ 
 E^{\ }_{\bm{k}}=
\sqrt{\varepsilon^{2}_{\bm{k}}+\Delta^{2}_{\bm{k}}}
$.
\end{subequations}
Both $\delta^{\ }_{0}$ and $\Delta^{\ }_{s}$ represent symmetry-breaking terms
that lower the symmetry from tetragonal to orthorhombic in an effective one-band
description of a single copper-oxygen plane.
The $s$-wave component $\Delta^{\ }_{s}$ is isotropic
[compared with the extended $s$-wave admixture from 
Eq.~(\ref{eq: extended s wave})]. 
The Raman vertices $\gamma^{\ }_{\bm{k}}$ can be
classified according to the irreducible representations of the
symmetry group of the crystal. For a crystal with tetragonal symmetry
(point group $D_{4h}$), the relevant symmetries are the $B^{\ }_{1g}$,
$B^{\ }_{2g}$, and $A^{\ }_{1g}$ polarizations.  
As we shall consider a model with subdominant orthorhombic distortions,
$\delta^{\ }_{0}\ll1$,
$\Delta^{\ }_{s}\ll\Delta^{\ }_{0}$,
we will use, in what follows,
the notation of tetragonal symmetry.\cite{remark1}
If so, we can identify the  
$B^{\ }_{1g}$, $B^{\ }_{2g}$, and $A^{\ }_{1g}$
channels for Raman scattering with the Raman vertices
\begin{subequations}
\label{eq:newvertices}
\begin{eqnarray}
\gamma^{\ }_{B^{\ }_{1g}\,\bm{k}}&\propto&
t 
\left[  
\left(1+\delta^{\ }_0\right)\cos k^{\ }_x 
- 
\left(1-\delta^{\ }_0\right)\cos k^{\ }_y 
\right],
  \\
\gamma^{\ }_{B^{\ }_{2g}\,\bm{k}}&\propto&
 4t'\sin k^{\ }_x\sin k^{\ }_y ,
\\
\gamma^{\ }_{A^{\ }_{1g}\,\bm{k}}&\propto&
 t\left[ 
(1+\delta^{\ }_0)\cos k^{\ }_x 
+ 
(1-\delta^{\ }_0)\cos k^{\ }_y 
\right]
\nonumber\\
&&
+
4t'\cos k^{\ }_x\cos k^{\ }_y,
\end{eqnarray}
\end{subequations}
respectively.

The electronic Raman response is calculated in the gauge invariant form
assuming that the quasiparticles interact through the long-range
Coulomb potential
$
V^{\ }_{\boldsymbol{q}}=
\frac{4\pi e^2}{\boldsymbol{q}^2}.
$
For the tetragonal symmetry, the RPA for the polarization-dependent 
Raman response function 
$\chi^{\ }_{\gamma}(\omega)$ 
was derived in Refs.~\onlinecite{Devereaux95,Klein84}
and \onlinecite{Monien90,Krantz95,Strohm97,Devereaux07}
and shown to be well defined in the limit $\boldsymbol{q}\to0$. 

Its generalization 
to orthorhombic symmetry leads to
\begin{subequations}
\label{eq: Raman responses B1g B2g A1g}
\begin{eqnarray}
\chi^{\ }_{B^{\ }_{1g}}(\omega)
&=&
\left\langle 
\gamma^{2}_{B^{\ }_{1g} } 
\theta^{\ }_{\bm{k}}
\right\rangle^{\ }_{\omega}
-
\frac{
\left\langle 
\gamma^{\ }_{B^{\ }_{1g}} 
\theta^{\ }_{\bm{k}}
\right\rangle^{2}_{\omega}
     }
     {
\left\langle 
\theta^{\ }_{\bm{k}}
\right\rangle_{\omega}
     },
\label{newb1g}
\\
\chi^{\ }_{B^{\ }_{2g}}(\omega)&=&
\left\langle 
\gamma^{2}_{B^{\ }_{2g}} 
\theta^{\ }_{\bm{k}}
\right\rangle^{\ }_{\omega},
\label{newb2g}
\\
\chi^{\ }_{A^{\ }_{1g}}(\omega)&=&
\left\langle 
\gamma^{2}_{A^{\ }_{1g}} 
\theta^{\ }_{\bm{k}}
\right\rangle^{\ }_{\omega}
-
\frac{
\left\langle 
\gamma^{\ }_{A^{\ }_{1g}}
\theta^{\ }_{\bm{k}}
\right\rangle^{2}_{\omega}
     }
     {
\left\langle 
\theta^{\ }_{\bm{k}}
\right\rangle_{\omega}
     },
\label{newa1g}
\end{eqnarray}
\end{subequations}
in the $B^{\ }_{1g}$, $B^{\ }_{2g}$, and $A^{\ }_{1g}$
channels for Raman scattering, respectively.
As usual, the bracket (in a box of volume $V$)
\begin{eqnarray} \label{eq: tsuneto function}
\left\langle (\cdots) \theta^{\ }_{\bm{k}} \right\rangle^{\ }_{\omega}=
&
\frac{1}{V}
\sum\limits_{\bm{k}}
 (\cdots)
\Delta^{2}_{\bm{k}}  \tanh\left(\frac{E^{\ }_{\bm{k}}}{2T}\right)
\\
&\times
\left(
\frac{1/E^{2}_{\bm{k}}}{\omega+i\eta+2E^{\ }_{\bm{k}}}
-
\frac{1/E^{2}_{\bm{k}}}{\omega+i\eta-2E^{\ }_{\bm{k}}}
\right)
\nonumber
\end{eqnarray}
denotes an average
over the BZ weighted by the Tsuneto function $\theta^{\ }_{\bm{k}}$.\cite{Tsuneto1960}
The second term in Eqs.~\re{newb1g} and~\re{newa1g} is commonly called
screening term. It can be viewed as originating from the Goldstone mode of the 
superconductor and ensures gauge invariance of the Raman response.\cite{Devereaux95}
For tetragonal symmetry only screening terms in the $A^{\ }_{1g}$ scattering channel 
are possible. That is, the screening term in Eq.~\re{newb1g} 
is vanishing identically if the SC quasiparticle dispersion is 
of the tetragonal symmetry. 
In the presence of orthorhombic distortions of the YBCO type
 screening terms can affect the $B^{\ }_{1g}$  
channels  on general symmetry grounds, but are absent in the $B^{\ }_{2g}$ channel.
Similarly, in orthorhombic Bi$_2$Sr$_2$CaCu$_2$O$_{8+x}$, whose crystallographic 
axes are rotated by 45$^\circ$ with respect to the CuO bonds,   screening terms 
arise in the $B^{\ }_{2g}$ channel, but do not affect the $B^{\ }_{1g}$ channel. 
In the following we shall only consider orthorhombic superconductors of the YBCO type
and disregard any possible screening terms in the $B^{\ }_{2g}$ channel. 
The presence of screening terms in the $B^{\ }_{1g}$ or $B^{\ }_{2g}$ 
channels for orthorhombic superconductors has been
previously reported in the literature, see Refs.~\onlinecite{Nemetschek98}
and \onlinecite{Strohm97}.

%
\begin{figure}[t!]
\includegraphics[width=0.47\textwidth]{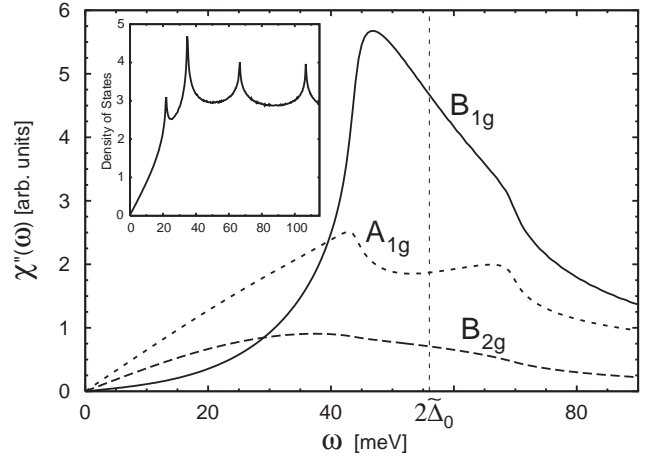}
\caption{
  Calculated electronic Raman response of $(d+s)$-wave
  superconductors for the $B^{\ }_{1g}$, $B^{\ }_{2g}$, 
  and $A^{\ }_{1g}$ channels (tetragonal notation), respectively,
  according to Eq.~(\ref{eq: Raman responses B1g B2g A1g}).
  The $B^{\ }_{1g}$ pair-breaking peak
  softens to $\omega \simeq 44\mbox{meV}<2\widetilde{\Delta}^{\ }_0  \simeq 56\mbox{meV}$
  where $\widetilde{\Delta}^{\ }_0$ is the SC gap maximum on the FS
  defined in Eq.~(\ref{eq: def 56 meV}). 
  The $A^{\ }_{1g}$ spectrum has been multiplied by the factor of $20$.
  Inset: ``bare'' density of states in the superconducting state for
  orthorhombic symmetry.
        }
\label{Fig: allpol}
\end{figure}
%

\section{Discussion} 
We have computed numerically the Raman response
(\ref{eq: Raman responses B1g B2g A1g})
using the parameters
$t= 200$~meV, 
$\delta^{\ }_0=-0.03$,
$t'/t=-0.4$,
$\mu/t = -1.2$,
$\Delta^{\ }_0 = 30$~meV, 
and
$\Delta^{\ }_s = 6$~meV.
We also choose the damping $\eta=1$~meV.
This choice yields a SC quasiparticle dispersion that
is in qualitative agreement with the one measured by photoemission experiments on
untwinned YBCO close to optimal doping.\cite{Schabel98}

In Fig.~\ref{Fig: allpol} we show the resulting Raman response for
$(d+s)$-wave superconductors. We find that the low-energy asymptotics,
$
\chi''_{B^{\ }_{1g}}(\omega)\propto
\left(\omega/2\widetilde{\Delta}^{\ }_0\right)^3
$,
$\chi''_{B^{\ }_{2g}}(\omega)\propto 
\left(\omega/2\widetilde{\Delta}^{\ }_0\right)
$, and
$\chi''_{A^{\ }_{1g}}(\omega)\propto
\left(\omega/2\widetilde{\Delta}^{\ }_0\right)$,
remain essentially unchanged 
with or without the symmetry-breaking terms $\delta^{\ }_{0}$ and $\Delta^{\ }_{s}$.
Here we have introduced the absolute maximum 
\begin{eqnarray}
\widetilde{\Delta}^{\ }_0=
\max_{{\bf k} \in \mathrm{FS}}\,\Delta({\bf k}) \simeq 28\,\mathrm{meV}
\label{eq: def 56 meV}
\end{eqnarray}
of the gap over the ``normal-state'' Fermi surface (FS) 
with full tetragonal symmetry ($\delta^{\ }_{0}=\Delta^{\ }_{s}=0$).
We also find that the relative peak heights in the $B^{\ }_{1g}$ and $A^{\ }_{1g}$ responses
 remain  \cite{Devereaux95} 
after switching on $\delta^{\ }_{0}>0$ and $\Delta^{\ }_{s}>0$.
These properties are robust to the lowering of the tetragonal to the 
orthorhombic symmetry as they owe their existence to that of the nodes
of the SC gap. There are nevertheless important
changes in the line shapes of the $A^{\ }_{1g}$ and $B^{\ }_{1g}$ responses due
to orthorhombicity which reflect the change 
in the underlying density of states (see inset of Fig.~\ref{Fig: allpol}).
Most importantly, the $B^{\ }_{1g}$ pair-breaking peak is not located any
longer at the absolute gap maximum $\omega=2\widetilde{\Delta}^{\ }_0 \simeq 56$~meV. 
Instead, it is shifted by $\simeq 2\Delta^{\ }_s$ to a lower value
due to the additional screening term in Eq.\ (\ref{newb1g})
and thus its position tracks the position of the double peaks in
the $A^{\ }_{1g}$ channel.\cite{footnoteA1g}
This is interesting
because it demonstrates that the common lore that the $B^{\ }_{1g}$
pair-breaking peak is always at $\omega=2\widetilde{\Delta}^{\ }_0$ is not correct in
$(d+s)$-wave superconductors. 
In passing, we note that a comparison between the $(XX)$-polarization channel, 
$\hat{e}^{O}=\hat{e}^{I}=\left(1\atop 0\right)$,
and $(YY)$-polarization channel,
$\hat{e}^{O}=\hat{e}^{I}=\left(0\atop 1 \right)$,
reveals only minor differences in the peak positions and line shapes 
(not shown).

Recently, it has been proposed that vertex corrections to the Raman response function
due to a spin-mediated pairing interaction can lead to the development of
a resonance below $2 \widetilde{\Delta}^{\ }_0$.%
\cite{Chubukov99,Venturini00,Zeyher02,Chubukov05}
Inclusion of these interaction effects would lead to an additional shift of the $B^{\ }_{1g}$
pair-breaking peak to lower frequencies. 
However, the magnitude of this shift depends on model assumptions and
it remains an open question, whether these vertex corrections
are of any relevance for the interpretation of the Raman data on
high-$T^{\ }_c$ cuprates. 

\begin{figure}[t!]
\includegraphics[width=0.47 \textwidth]{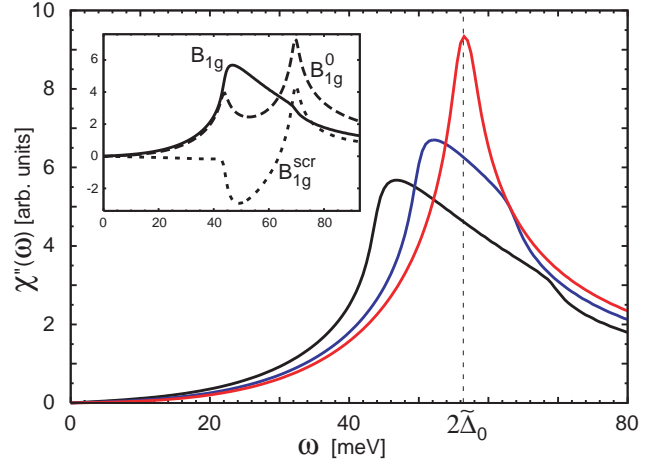}
\caption{(color online) 
  Calculated electronic Raman response of $(d+s)$-wave
  superconductors for the $B^{\ }_{1g}$
  channel (tetragonal notation)
  according to Eq.~(\ref{eq: Raman responses B1g B2g A1g}).
  The fermiology parameters from Eq.~(\ref{eq: master equations}) are
  the same as in Fig.~\ref{Fig: allpol} except for
  $\Delta^{\ }_s = 0$~meV (red), 
  $\Delta^{\ }_s = 3$~meV (blue), 
  and $\Delta^{\ }_s = 6$~meV (black).
  Inset: decomposition of the left-hand side of Eq.~(\ref{newb1g})
  into the sum of the bare contribution
  $B^{0}_{1g}$
  and the screening contribution
  $B^{\mathrm{scr}}_{1g}$
  on the right-hand side of Eq.~(\ref{newb1g})
  when $\Delta^{\ }_s = 6$~meV.
        }
\label{Fig: shift}
\end{figure}

In order to analyze in more detail the softening of the $B^{\ }_{1g}$
pair-breaking peak due to the screening term, we show
in Fig.~\ref{Fig: shift} the results for $\Delta^{\ }_s=0$~meV, $3$~meV, and
$6$~meV, respectively, with all other parameters held fixed. 
We find that increasing $\Delta^{\ }_s$ induces
a reduction of the peak intensity
and a shift to lower frequencies of the peak positions.
A closer inspection of Eq.~(\ref{newb1g}) reveals that the bare contribution
\begin{equation}
B^{0}_{1g}(\omega)=
\mathrm{Im} \left\langle
\gamma^{2}_{B^{\ }_{1g} }
\theta^{\ }_{\bm{k}}
\right\rangle^{\ }_{\omega}
\label{eq: bare}
\end{equation}
exhibits two peaks located at
$\omega=2(\widetilde{\Delta}^{\ }_0\mp\Delta^{\ }_s)$,
respectively.
This double peak structure  
is a reflection of the bare quasiparticle density of states 
that features two local maxima (inset of Fig.~\ref{Fig: allpol})
at half the frequencies of the peaks in~$B^{0}_{1g}$.
In contrast, the screening contribution
\begin{eqnarray}
B^{\mathrm{scr}}_{1g}(\omega)=
\mathrm{Im} \left[
\left\langle
\gamma^{\ }_{B^{\ }_{1g} }
\theta^{\ }_{\bm{k}}
\right\rangle^{2}_{\omega}
/
\left\langle
\theta^{\ }_{\bm{k}}
\right\rangle_{\omega}
\right] ,
\label{eq: scr}
\end{eqnarray}
shows a broad dip followed by a peak
(see the inset of Fig.~\ref{Fig: shift}).
That is, an enhancement
of the scattering efficiency by screening 
(\emph{antiscreening}) occurs
at low frequencies $\omega \lesssim 2 \widetilde{\Delta}_0$,
whereas at high frequencies $\omega \gtrsim 2 \widetilde{\Delta}_0$
 the scattering efficiency is reduced by the screening term.
 To show how antiscreening arises,
we rewrite the imaginary part of the $B^{\ }_{1g}$ screening term  as
\begin{eqnarray} \label{eq: b1g screening term}
B^{\mathrm{scr}}_{1g}(\omega) 
&=&
\frac{
2 \left\langle \theta^{\prime}_{\bm{k}}  \right\rangle_{\omega} 
\left\langle \gamma^{\ }_{B^{\ }_{1g} }  \theta^{\prime}_{\bm{k}}  \right\rangle_{\omega}
\left\langle \gamma^{\ }_{B^{\ }_{1g} }  \theta^{\prime \prime}_{\bm{k}}  \right\rangle_{\omega}
}
{
\left\langle \theta^{\prime}_{\bm{k}} \right\rangle^2_{\omega}
+
\left\langle \theta^{\prime \prime}_{\bm{k}} \right\rangle^2_{\omega}
}
\\
& &\qquad
+
\frac{
\left\langle  \theta^{\prime \prime}_{\bm{k}} \right\rangle_{\omega}
\left[
\left\langle 
\gamma^{\ }_{B^{\ }_{1g}}  \theta^{\prime \prime }_{\bm{k}}  
\right\rangle^2_{\omega}
-
\left\langle \gamma^{\ }_{B^{\ }_{1g}}  \theta^{\prime }_{\bm{k}}  \right\rangle^2_{\omega}
\right]
}
{
\left\langle \theta^{\prime}_{\bm{k}} \right\rangle^2_{\omega}
+
\left\langle \theta^{\prime \prime}_{\bm{k}} \right\rangle^2_{\omega}
},
\nonumber
\end{eqnarray}
where the real and imaginary parts of the Tsuneto functions are denoted by 
$\theta^{\prime}_{\bm{k}}$
and $\theta^{\prime \prime }_{\bm{k}}$, respectively.
It is instructive to approximate the quasiparticle dispersion by
\begin{subequations}
\begin{equation}
E^{2}_{ k, \phi } \simeq \varepsilon^2_{k} +  | \Delta^{\ }_{\phi} |^2
\end{equation}
with the linearized dispersion 
$ \varepsilon^{\ }_{k}= v^{\ }_F ( k -k^{\ }_F ) $
and the gap function 
$ \Delta^{\ }_{\phi} =  \widetilde{\Delta}^{\ }_0 \cos(2 \phi) + \Delta^{\ }_s $
that depends only on the Fermi surface angle $\phi$. 
Similarly, we replace the $B^{\ }_{1g}$ vertex by its 
angular-dependent form,
\begin{equation}
\gamma^{\ }_{B^{\ }_{1g}} \simeq \cos(2 \phi).
\end{equation}
\end{subequations}
With these simplifications and in the zero-temperature limit the
imaginary part of the BZ averages reduce to
\begin{eqnarray} \label{eq: approximate tsuneto fun}
\left\langle  ( \cdots)  \theta^{\prime \prime  }_{\bm{k}} \right\rangle
& \simeq &
  \int \frac{ k dk}{ 2 \pi}
\int \frac{ d \phi}{2 \pi}
( \cdots )
 \frac{ \Delta^2_{\phi } }{ E^2_{k, \phi}    } 
\delta \left( \omega - 2 E^{\ }_{k, \phi} \right)
 \nonumber\\
&\simeq&
  \int_{- \delta k}^{+\delta k}  \frac{ d \tilde{k} }{ 2 \pi}
\sum_{\phi^{\ }_i} ( \cdots )
\frac{ \tilde{k} + k^{\ }_F }{2 E^{\ }_{\tilde{k}+k^{\ }_F, \phi^{\ }_i}} 
\frac{ | \Delta^{\ }_{\phi^{\ }_i} | }{ | \Delta^{\prime}_{\phi^{\ }_i} | },
\end{eqnarray}
where we have restricted the BZ integration
to a ring $- \delta k < k -k^{\ }_F < \delta k$ around the Fermi surface, 
the summation runs over all 
$\phi^{\ }_i  \in \left\{ \phi \; | \; \omega = 2 E^{\ }_{k,\phi} \right\}$,
and $\Delta^{\prime}_{\phi}$ denotes
the derivative of the gap  function $\Delta^{\ }_{\phi}$.
The imaginary part of the averaged Tsuneto function 
$\langle \theta^{\prime \prime}_{\bm{k}} \rangle$
 is a positive function of $\omega$ and exhibits a positive divergence whenever 
$ \Delta^{\prime}_{\phi^{\ }_i} = 0$,
 i.e., at the frequencies
 \begin{eqnarray}
 \begin{split}
 \omega^{(1)}_c \simeq E^{\ }_{k^{\ }_F, \frac{\pi}{2} } 
&=  2 (\widetilde{\Delta}_0 - \Delta^{\ }_s ),
\\
  \omega^{(2)}_c \simeq E^{\ }_{k^{\ }_F, 0} 
&= 2 (\widetilde{\Delta}_0 + \Delta^{\ }_s ),
\end{split}
 \end{eqnarray}
as can be seen from the last line of Eq.~\re{eq: approximate tsuneto fun}.
On the other hand,  $\langle \gamma^{\ }_{B^{\ }_{1g}}  \theta^{\prime \prime} \rangle$
possesses  a negative divergence at $\omega^{(1)}_c$ and a positive
divergence at $\omega^{(2)}_c$, and changes
its sign at $\sim 2 \widetilde{\Delta}_0$, since
the $B^{\ }_{1g}$ vertex $\gamma^{\ }_{B^{\ }_{1G}} \simeq \cos{2 \phi}$
exhibits a sign change along the Fermi surface.
Any divergence in the frequency dependence of
$\langle \theta^{\prime \prime} \rangle$ and
$\langle \gamma^{\ }_{B^{\ }_{1g}} \theta^{\prime \prime} \rangle$
results in a steplike discontinuity in $\langle \theta^{\prime} \rangle$
and $\langle \gamma^{\ }_{B^{\ }_{1g}} \theta^{\prime} \rangle$, respectively,
due to the Kramers-Kronig relation. Vice versa, from the absence
of any steplike discontinuity in the imaginary part of the BZ averages
follows the absence of any divergence in the real part of the
corresponding quantity.
Furthermore,
it turns out that both $\langle \theta^{\prime} \rangle$ 
and $\langle \gamma^{\ }_{B^{\ }_{1g}} \theta^{\prime} \rangle$ are
positive in the frequency range 
$0 \leq \omega \lesssim 2 ( \widetilde{\Delta}_0 + \Delta^{\ }_s )$.
Taking all these observations together, we find
that the first term in the $B^{\ }_{1g}$ screening function \re{eq: b1g screening term}
features a negative divergence at $\omega = 2 ( \widetilde{\Delta}_0 - \Delta^{\ }_s )$, which
is compensated by a positive divergence of the second term.
But then, at $\omega = 2 ( \widetilde{\Delta}_0 + \Delta^{\ }_s )$ 
both terms in Eq.~\re{eq: b1g screening term}
show a positive divergence, which results in a strong screening of the second peak in 
$B^0_{1g}$.
At $\omega \simeq 2   \widetilde{\Delta}_0$ the first term in Eq.~\re{eq: b1g screening term}
is vanishing, whereas the second term is negative, which yields to antiscreening 
of the $B^{\ }_{1g}$   scattering efficiency.

It is important to note that the effects induced by
a nonvanishing isotropic $s$-wave gap $\Delta^{\ }_{s}$
in Fig.~\ref{Fig: shift}
would not occur had we assumed a
subdominant extended $s$-wave gap such as
\begin{equation}
\begin{split}
\Delta^{\ }_{\bm{k}}=
&\Delta^{\ }_{0} 
\left(
\cos k^{\ }_x 
-
 \cos k^{\ }_y
\right) / 2 
\\
&+
\Delta^{\mathrm{ext}}_{s} 
\left(
\cos k^{\ }_x 
+
 \cos k^{\ }_y
\right) / 2 .
\label{eq: extended s wave}
\end{split}
\end{equation}
Contrary to the isotropic subdominant $s$-wave admixture
$\Delta^{\ }_{s}$ in Eq.\ (\ref{eq: def isotropic s wave gap}),
the extended $s$-wave admixture in Eq.\ \re{eq: extended s wave}   
only leads to a shift of the nodal line, 
but does not give a different  absolute gap value
at $(\pi, 0)$ compared to $(0, \pi)$ (see Fig.~2(d) in Ref.~\onlinecite{Schnyder06}).
From the above discussion one infers
that the property 
$|\Delta^{\ }_{\phi= \frac{\pi}{2}} | \ne | \Delta^{\ }_{\phi=0} | $ 
is crucial for the splitting of the $B^{\ }_{1g}$ pair-breaking peak 
as well as for the appearance of antiscreening in the $B^{\ }_{1g}$ channel.
Thus, we conclude that to a first approximation an extended
$s$-wave admixture leaves the $B^{\ }_{1g}$ response unchanged,
an expectation that we have confirmed by numerical calculations.

\section{Superconductivity-induced changes in phonon line shapes}

In principle, the existence of  a subdominant and isotropic $s$-wave component 
$\Delta^{\ }_{s}$
to the $d$-wave SC gap can be extracted from a line-shape analysis
of the $A^{\ }_{1g}$ and $B^{\ }_{1g}$ electronic responses in a
polarization-resolved Raman-scattering experiment.
However, as part of the electronic Raman signal is in general masked 
by phonon excitations, it might be hard to detect experimentally these changes in
the line shape. On the other hand, we argue in the following that 
polarization-resolved Raman scattering on
phonons can be used to extract a
signature of  $\Delta^{\ }_{s}$. 

\begin{figure}[t!]
\includegraphics[width=0.47\textwidth]{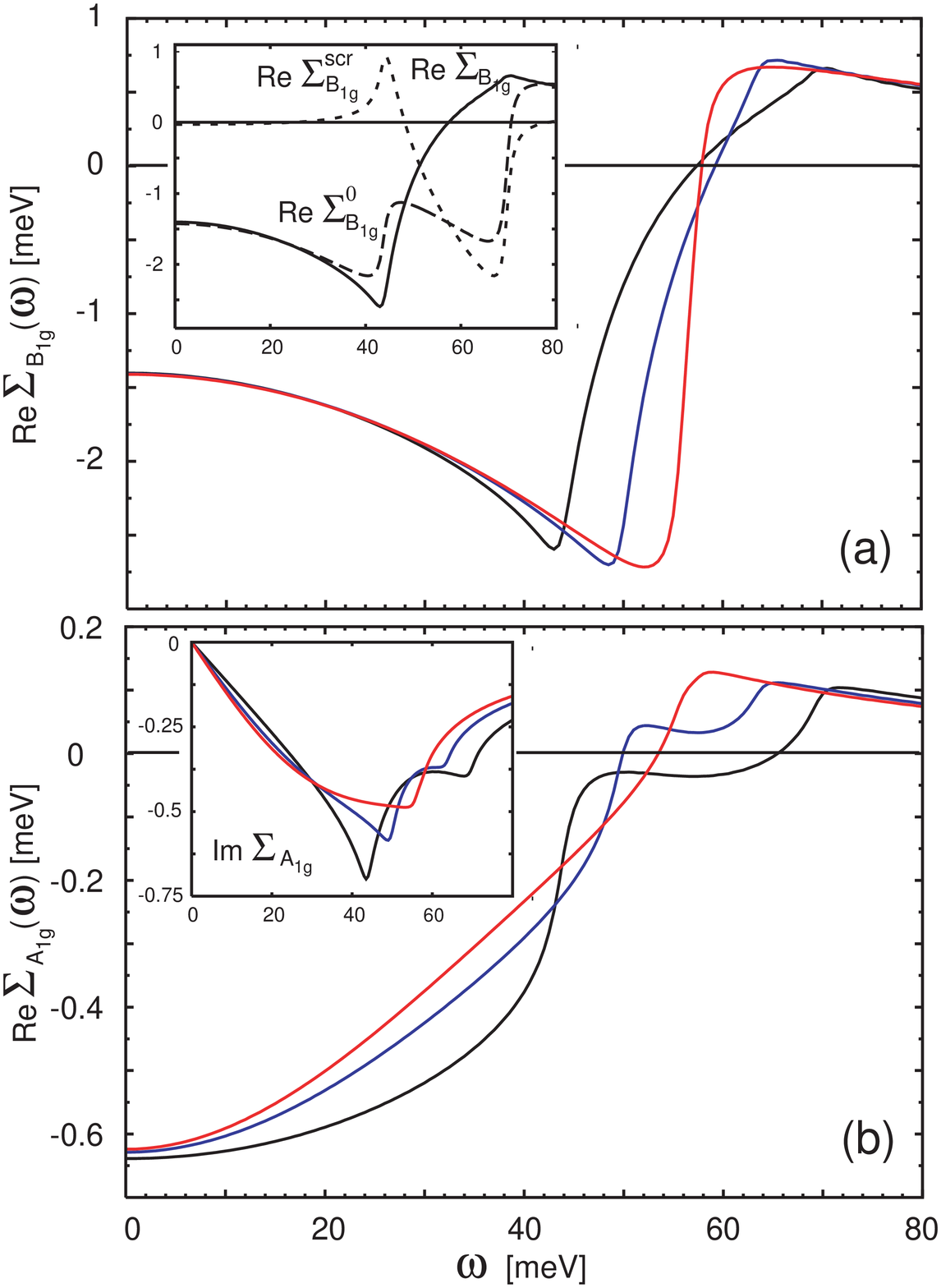}
\caption{(color online)
Panel (a) displays the
 real part of the superconductivity-induced phonon self-energy 
$\mathrm{Re} \Sigma^{\ }_{B^{\ }_{1g}} (\omega)$ 
 of the $B^{\ }_{1g}$ phonon with $g^2_{B^{\ }_{1g}} = 0.1 \Delta^{\ }_0 / N_F$ and
 at zero temperature according to Eq.~\re{eq: phonon self energy}.
  The fermiology parameters are the same as in Fig.~1
 except for $\Delta^{\ }_s$, which varies
 form $\Delta^{\ }_s = 0$ meV (red), to $\Delta^{\ }_s = 3$ meV (blue),
 and $\Delta^{\ }_s = 6$ meV (black). The inset shows
 the decomposition of the left-hand side of 
 Eq.~\re{eq: phonon self energy} into a sum of the bare
 contribution $\mathrm{Re} \Sigma^{0}_{B^{\ }_{1g}}$
 and the screening contribution 
 $\mathrm{Re} \Sigma^{\mathrm{scr} }_{B^{\ }_{1g}}$
 when $\Delta^{\ }_s = 6$ meV.
 Panel (b) shows   the real and imaginary parts
 of the superconductivity-induced self-energy of the 
 $A^{\ }_{1g}$ phonon with  $g^2_{B^{\ }_{1g}} = 1.0 \Delta^{\ }_0 / N_F$
 and the same fermiology parameters as in panel (a).
 Negative (positive) values correspond to
 softening (hardening) in the main frame
 and broadening (sharpening) in the inset of panel (b).
Here  $N_F$ denotes the ``normal-state'' density of states per spin at the Fermi energy.
       }
\label{fig: phonon}
\end{figure}

To substantiate this point, we calculate the superconductivity-induced
changes in the self-energy of optical, zone-center ($\bm{q} = 0$) phonons.
Thereto, we consider a linear coupling between
electrons and phonons 
\begin{equation} 
H^{\ }_{\mathrm{el-ph}}= 
\frac{1}{V}
\sum_{\bm{k},\bm{q},\gamma,\sigma}\, 
g^{\gamma     }_{\bm{k}       ,\bm{q}}\, 
c^{\dag  }_{\bm{k}+\bm{q},\sigma} 
c^{\     }_{\bm{k}       ,\sigma}
\left( 
b^{\   }_{ \bm{q}, \gamma}
+
b^{\dag}_{-\bm{q}, \gamma} 
\right),
\label{eq: electron-phonon interaction}
\end{equation}
where $b^{\dag}_{\bm{q},\gamma}$ and $b^{\   }_{\bm{q},\gamma}$ are the
creation and annihilation operators of phonons in a given branch $\gamma$
with phonon frequency $\omega_{\gamma}$,
respectively, and $g^{\gamma}_{\bm{k},\bm{q}}$ denotes the electron-phonon
coupling. 
The form of the electron-phonon interaction~\re{eq: electron-phonon interaction}
is the one for the nonresonant electronic Raman scattering provided
the effective Raman vertex is replaced by the electron-phonon vertex
$g^{\gamma }_{\bm{k},\bm{q}}$. Hence, within an RPA treatment of the Coulomb 
interactions, we find that in the $\bm{q} \to 0$ 
limit the superconductivity-induced changes
in the phonon self-energy are given by 
\cite{Zeyher90,Nicol93,Devereaux94}
\begin{eqnarray} \label{eq: phonon self energy}
\Sigma^{\ }_{\gamma} (\omega)= 
- 
\left\langle 
\left( 
g^{\gamma}_{\bm{k},0}  
\right)^2 
\theta_{\bm{k}} 
\right\rangle^{\ }_{\omega} 
+ 
\frac{
\left\langle 
g^{\gamma}_{\bm{k},0} 
\theta_{\bm{k}}
\right\rangle^{2}_{\omega} 
     }
     {
\left\langle 
\theta_{\bm{k}} 
\right\rangle_{\omega} 
     },
\end{eqnarray}
where the angular brackets are defined by 
Eq.~\re{eq: tsuneto function}.
The symmetry of the optical phonons 
is encoded in the matrix element $g^{\gamma}_{\bm{k},0} $.
The electron-phonon coupling of the 
$A^{\ }_{1g}$ and $B^{\ }_{1g}$ phonons 
are in a first approximation given by
\begin{equation}
\begin{split}
g^{B^{\ }_{1g} }_{\bm{k},0}
&=
g^{\ }_{B^{\ }_{1g}} ( \cos k^{\ }_x - \cos k^{\ }_y ) / 2,
\\
g^{A^{\ }_{1g} }_{\bm{k},0}
&=
g^{\ }_{A^{\ }_{1g}} ( \cos k^{\ }_x + \cos k^{\ }_y ) / 2.
\end{split}
\end{equation}
with the electron-phonon coupling constants $g^{\ }_{B^{\ }_{1g}}$
and $g^{\ }_{A^{\ }_{1g}}$.
In Fig.~\ref{fig: phonon} we have
numerically evaluated the superconductivity-induced changes
in the self-energy $\Sigma^{\ }_{\gamma} (\omega)$
for  $B^{\ }_{1g}$ and $A^{\ }_{1g}$ phonons in a
 $(d+s)$-wave superconductor.
To estimate the size of these effects we have assumed 
some typical values for the phonon coupling strength $g^{\ }_{\gamma}$
that are in rough overall agreement with the observed  Raman shifts
in YBCO.\cite{Thomsen90,Friedl90,Limonov98,Bock99,Hewitt04}
A negative (positive) $\mbox{Re}\,\Sigma^{\ }_{\gamma} (\omega)$
corresponds to a softening (hardening) 
of the  phonon below $T^{\ }_c$,
whereas a negative (positive) $\mbox{Im}\,\Sigma^{\ }_{\gamma} (\omega)$ 
leads to a broadening (sharpening) of the phonon linewidth below $T^{\ }_c$.

For $B^{\ }_{1g}$ phonons [see Fig.\ \ref{fig: phonon}(a)]
we find that the real part of the phonon self-energy 
crosses zero around  $2 \widetilde{\Delta}_0$ irrespective of
the value of $\Delta^{\ }_s$. However, upon inclusion
of a subdominant $s$-wave gap the maxima and
minima of $\mathrm{Re} \Sigma^{\ }_{B^{\ }_{1g}}$ are shifted
to $2( \widetilde{\Delta}_0 + \Delta^{\ }_s)$ and
$2( \widetilde{\Delta}_0 - \Delta^{\ }_s)$, respectively.
Hence, the maximal softening of the phonon frequency
occurs at a frequency $2 \Delta^{\ }_s$ smaller than 
the gap maximum $\widetilde{\Delta}_0$.
The inset of Fig.~\ref{fig: phonon}(a) displays 
the decomposition of the real part of the self-energy  
$\mbox{Re}\,\Sigma^{\ }_{B^{\ }_{1g}} (\omega)$
into its screened $\mbox{Re}\,\Sigma^{\mathrm{scr}}_{B^{\ }_{1g}} (\omega)$
and bare parts $\mbox{Re}\,\Sigma^{0}_{B^{\ }_{1g}} (\omega)$.
The frequency dependence of these two contributions
mimics that  of $B^0_{1g}(\omega)$ and $B^{\mathrm{scr}}_{1g}(\omega)$, which
we discussed in the previous Section.
A finite $s$-wave component splits the step in
$\mbox{Re}\,\Sigma^{0}_{B^{\ }_{1g}} (\omega)$
into two steps located at  $2 ( \widetilde{\Delta}_0 \mp \Delta^{\ }_s )$ .
 In between these two steps the screening contribution
 $B^{\mathrm{scr}}_{1g}(\omega)$ is negative, which leads
 to antiscreening.
Neglecting the influence of  a nonzero $\delta_0$, 
one can infer the Kramers-Kronig transform
of the curves in Fig.~\ref{fig: phonon}(a), 
directly from Fig.~\ref{Fig: shift} and thereby obtain
the imaginary part of the self-energy $\mbox{Im}\,\Sigma^{\ }_{B^{\ }_{1g}} (\omega)$.
Due to the $s$-wave admixture, the frequency where
the maximal broadening occurs is shifted from 
$2 \widetilde{\Delta}^{\ }_0$ to
$  2 ( \widetilde{\Delta}^{\ }_0 - \Delta^{\ }_s )$.\cite{Friedl90}

Similar effects occur for the self-energy of $A^{\ }_{1g}$ phonons 
[see Fig.~\ref{fig: phonon}(b)].
An isotropic $s$-wave component splits the step in the real part 
$\mbox{Re}\,\Sigma^{\ }_{A^{\ }_{1g}} (\omega)$
into two steps, thereby shifting the crossing point of 
$\mbox{Re}\,\Sigma^{\ }_{A^{\ }_{1g}} (\omega)$
with the zero line to lower frequencies ($\Delta_s = 3$ meV)
or to higher frequencies ($\Delta_s = 6$ meV).
Contrary to $B^{\ }_{1g}$ phonons, antiscreening effects are absent in the self-energy
of $A^{\ }_{1g}$ phonons.

Finally, we note that the effects on the superconductivity-induced changes
of the phonon self-energy induced by an isotropic $s$-wave gap
are again absent for a gap of pure
$d$-wave character or for an extended $s$-wave admixture such as in
Eq.~(\ref{eq: extended s wave}). 
The effects on the phonon self-energy reported in this paper 
therefore constitute  a fingerprint  
of an isotropic $s$-wave admixture to the pairing symmetry.

\section{Conclusions} 
We have calculated the polarization-dependent Raman
response for electrons and phonons in $(d+s)$-wave superconductors. We
find that the screening terms in the $B^{\ }_{1g}$ channel lead to a softening of
the $B^{\ }_{1g}$ pair-breaking peak of the order of $2\Delta^{\ }_s$, 
i.e., twice the value of the isotropic $s$-wave component to the SC gap.
This fact calls into question the long-standing interpretation that 
the $B^{\ }_{1g}$ pair-breaking peak is located at
$\omega=2\widetilde{\Delta}^{\ }_0$ and thus can be directly
used as a measure of the $d$-wave component
of the SC gap when orthorhombicity is present. 
Secondly, we have estimated the effects of 
a subdominant $s$-wave admixture on the 
superconductivity-induced phonon renormalizations. 
These effects, although comparatively small, 
might serve as a fingerprint
of an isotropic $s$-wave admixture to the pairing symmetry.
\acknowledgments
This work was supported by the Swiss National Science Foundation under Grant
No. 200021-101765/1. 
We thank D.~Einzel for pointing out an alternative derivation of
Eq.~\re{eq: Raman responses B1g B2g A1g}, 
and are thankful to B.~Keimer, C.~Ulrich, M.~Bakr, 
M.~Sigrist, and M.\ Cardona for stimulating discussions.
D.M. thanks the ETH Z\"urich for hospitality and
gratefully acknowledges financial support from the Alexander von
Humboldt foundation.


\begin{thebibliography}{99}
\bibitem{remark1} 
If the crystal is orthorhombic ($D_{2h}$ point group), 
the $A^{\ }_{1g}$ and $B^{\ }_{1g}$ representations of the
tetragonal space group ($D_{4h}$ point group) become identical.
To distinguish between these two Raman tensors  
an extra parameter is needed. Here, we are resolving
this ambiguity by defining the $A^{\ }_{1g}$ and $B^{\ }_{1g}$
tensors in terms of the incoming and outgoing
light polarization vectors as follows:
We define $\gamma^{\ }_{B^{\ }_{1g}}$ by
$\gamma^{\ }_{(\hat{e}^{O}, \hat{e}^{I} )} = 
\gamma^{\ }_{( \hat{x}^{\prime} , \hat{y }^{\prime}) }$
and for $\gamma^{\ }_{A^{\ }_{1g}}$
we choose  
$\gamma^{\ }_{( \hat{e}^{O}, \hat{e}^{I} )} = 
\gamma^{\ }_{(\hat{x}^{\prime} , \hat{x}^{\prime} )} -  \gamma^{\ }_{(\hat{x}, \hat{y})}
$,
with the vectors
$\hat{x} =   \left(+1\atop 0 \right) $, 
$\hat{y} =   \left(0 \atop +1 \right) $, 
$\hat{x}^{\prime}  =  \frac{1}{\sqrt{2}}  \left(+1\atop +1 \right) $, 
and
$\hat{y}^{\prime}  =  \frac{1}{\sqrt{2}}  \left(- 1\atop +1 \right) $.
\end{thebibliography}
\end{document}